\documentstyle[12pt]{article}
\input{epsf.sty}

\def\a{\alpha}
\def\b{\beta}

\def\d{\delta}
\def\e{\epsilon}   
\def\f{\varphi}    
\def\g{\gamma}

\def\k{\kappa}     
\def\l{\lambda}
\def\m{\mu}

\def\p{\pi}       
\def\q{\theta}    

\def\D{\Delta}

\def\Q{\Theta}

\def\cF{{\cal F}}
\def\cG{{\cal G}}

\def\mh{\hat{\m}}

\def\Db{\D^{(0)}}
\newcommand{\ncm}{\newcommand}
\ncm{\rencm}{\renewcommand}
\ncm{\dsp}{\displaystyle}
\ncm{\nn}{\nonumber}
\ncm{\nnn}{\nonumber\\}
\ncm{\nit}{\noindent}
\ncm{\del}{\partial}
\ncm{\av}[1]{\mbox{$\langle #1 \rangle$}}
\ncm{\avc}[1]{\mbox{$\langle #1 \rangle_{\psi}$}}
\ncm{\half}{\mbox{{\small $\frac{1}{2}$}} }
\ncm{\quart}{\mbox{{\small $\frac{1}{4}$}} }
\ncm{\tq}{\mbox{{\small $\frac{3}{4}$}} }
\ncm{\third}{\mbox{{\small $\frac{1}{3}$}} }
\ncm{\sixth}{\mbox{{\small $\frac{1}{6}$}} }
\ncm{\eigth}{\mbox{{\small $\frac{1}{8}$}} }
\ncm{\thrhalf}{\mbox{{\small $\frac{3}{2}$}} }
\ncm{\thrfor}{\mbox{{\small $\frac{3}{4}$}} }
\ncm{\twothi}{\mbox{{\small $\frac{2}{3}$}} }
\ncm{\fivtwo}{\mbox{{\small $\frac{5}{2}$}} }
\ncm{\ninhalf}{\mbox{{\small $\frac{9}{2}$}} }
\ncm{\ninth}{\mbox{{\small $\frac{1}{9}$}} }
\ncm{\nist}{\mbox{{\small $\frac{9}{16}$}} }
\ncm{\df}{\mbox{$\partial_{\phi}$}}
\ncm{\dft}{\mbox{$\partial_{\phi}^2$}}
\ncm{\da}{\mbox{$\partial_{a}$}}
\ncm{\dat}{\mbox{$\partial_{a}^2$}}
\ncm{\dath}{\mbox{$\partial_{a}^3$}}
\ncm{\dx}{\mbox{$\partial_{x}$}}
\ncm{\dxi}{\mbox{$\partial_{x_i}$}}
\ncm{\dxt}{\mbox{$\partial^2_{x}$}}
\ncm{\dxit}{\mbox{$\partial^2_{x_i}$}}
\ncm{\dt}{\mbox{$\partial_{t}$}}
\ncm{\dtt}{\mbox{$\partial_{t}^2$}}
\ncm{\pf}{\mbox{$p_{\phi}$}}
\ncm{\Sh}{S}
\ncm{\un}{1\!\!1}
\ncm{\RE}{\mbox{Re}}
\ncm{\IM}{\mbox{Im}}
\ncm{\Tr}{\mbox{tr}\,}
\ncm{\diag}{\mbox{diag}\,}
\ncm{\Det}{\mbox{Det}\,}
\ncm{\ra}{\rightarrow}
\ncm{\la}{\leftarrow}
\ncm{\dg}{\dagger}
\ncm{\pr}{\prime}
\ncm{\ha}{\hat{a}}
\ncm{\hP}{\hat{P}}
\ncm{\sL}{\sqrt{\Lambda}}
\ncm{\lb}{\overline{\lambda}}
\ncm{\aldot}{\mbox{$\dot{\alpha}$}}
\ncm{\dota}{\mbox{$\dot{a}$}}
\ncm{\dotf}{\mbox{$\dot{\phi}$}}
\ncm{\dfo}{\mbox{$\partial_{\phi_0}$}}
\ncm{\aplt}{ \mbox{}_{\textstyle \sim}^{\textstyle < }     }
\ncm{\apgt}{ \mbox{}_{\textstyle \sim}^{\textstyle > }     }
\ncm{\Oa}{\mbox{$\mbox{O}(a)$}}
\ncm{\Sp}{\hspace{1.0cm}}
\def\be{\begin{equation}}
\def\ee{\end{equation}}
\def\bea{\begin{eqnarray}}
\def\eea{\end{eqnarray}}
\rencm{\thefootnote}{\mbox{\protect{$\fnsymbol{footnote}$}} }
\ncm{\front}[5]{
\begin{titlepage}
\noindent {#1} \hfill {#2}\\
\begin{center}
\vspace{1.5\baselineskip}
{\Large\bf  #3  } \\
\vspace{2\baselineskip}
\vspace{1.5\baselineskip}
 #4\\
\vspace{1.5\baselineskip}

University of California at San Diego,\\
Department of Physics 0319,
La Jolla, CA 92093, USA.

\end{center}
\vfill
{\bf Abstract}\\
 #5
\end{titlepage} }

\begin{document}
\front{September 1993}{UCSD/PTH 93-29}
{Almost Gauge invariant lattice actions for chiral gauge theories, using
Laplacian gauge fixing}
{Jeroen C. Vink\footnote{e-mail: vink@yukawa.ucsd.edu} }
{It is described how to obtain an almost gauge invariant lattice action $S$
for chiral gauge theories, or
other models in which a straightforward discretization leads to a
lattice action $S^\pr$ in which gauge invariance is broken.
The lattice action is `almost' gauge invariant, because  a local
gauge transformation leaves the action the same, up to a global gauge
transformation.
In this approach the action $S$ for
all gauge fields on a gauge orbit is the same as that of the action
$S^\pr$ evaluated for the gauge field fixed to a smooth gauge.
To define $S$ unambiguously, it must be possible to compute the gauge
fixed field unambiguously. This rules out gauge conditions which suffer
from Gribov ambiguities but it can be achieved by using the recently
proposed Laplacian gauge. When using the almost gauge invariant action $S$
in a numerical simulation, it is not necessary to fix the gauge, and hence
gauge fixing terms and  Fadeev-Popov ghosts are not required.
A hybrid Monte Carlo algorithm for simulations with this new
action is described and tested on a simple toy model. }

\section{Introduction}

When attempting to formulate chiral gauge theories using the lattice
as a regulator, it is found that the standard approaches lead to lattice
models in which the gauge invariance is broken by the lattice fermions.
E.g. with Wilson fermions the mass
term that removes the doublers breaks gauge invariance and with staggered
fermions only certain discrete transformations remain a symmetry of the
action.
Most approaches to deal with this breaking of gauge invariance
can be divided in two groups: those that impose gauge fixing
and those that avoid gauge fixing but restore gauge invariance with the
aid of an extra scalar field; see ref. \cite{Petc93} for a recent review.

In the non-gauge fixing approaches, the gauge degrees of freedom
corresponding to gauge transformations (`longitudinal modes')
appear as a strongly coupled (group-valued) scalar field in the
action.
One hopes, that in the extended fermion-scalar-gauge model there exists
a critical region in which the scalar field decouples such that
model recovers the desired symmetry
properties and particle content of the continuum target model.
It has been found that this scenario works for certains models in which
the gauge symmetry breaking is sufficiently small, e.g. in the massive
Yang-Mills model,
but it appears not to work in lattice approaches to chiral gauge theories,
using Wilson fermions \cite{SmSw92}, staggered fermions \cite{BoSm93} or
domain wall fermions \cite{GoJa93}.

The alternative approach uses gauge fixing. The underlying idea here
is that by  gauge fixing the continuum model, one can restrict the
saddle point configurations of the path integral to be smooth, as in
models without gauge symmetry, such that  also the integration over
quantum fluctuations in the path integral poses no particular difficulties.
Or putting it differently, the troublesome `longitudinal modes' of the
gauge field are removed by fixing the gauge.
At least in perturbation theory, it should be possible to recover
the BRST invariance of the continuum target model by adding a proper
set of counter terms \cite{Rome}.
To implement this nonperturbatively  on the lattice, one must discretize the
continuum action, including the gauge fixing  and ghost terms. However,
the Fadeev-Popov determinant that results after integrating out the ghosts
is not positive definite, because of the Gribov copies
\cite{Grib78}, and it is not clear how to handle such an action in a
numerical simulation.

In this letter, a different approach is explained, in which each gauge
field on a gauge orbit has the same action as the gauge field which
is fixed to a smooth gauge. This makes the action gauge invariant, up to
a global gauge transformation. However,
to make this definition unambiguous, it must be possible to compute the
smooth gauge section without Gribov ambiguities.
As in the non-gauge fixing approach, it is not necessary to
add gauge fixing terms or Fadeev-Popov ghosts to this action in
a numerical simulation.
We describe an (exact) algorithm for numerical simulations with this
almost gauge invariant action, that
combines the hybrid Monte Carlo method with a gauge
fixing prescription which is free from Gribov ambiguities.
In this letter we focus on the generic features of our method, without
reference to a specific model, but as is the case with the Rome
approach \cite{Rome}, our method could be applied to several of the
existing proposals for chiral fermions
on the lattice reviewed in ref. \cite{Petc93}.

\section{Making a gauge variant action almost gauge invariant}

Suppose we have discretized an action which is gauge invariant in the
continuum, but could not preserve the gauge invariance on the lattice.
Starting from a continuum action $S_{cont}$, we have obtained
a lattice action $S^\pr$, which reduces to $S_{cont}$ in the classical
continuum limit. I.e. for smooth classical fields,
$S^\pr = S_{cont}+O(a)$, with $a$ the lattice distance.
However, the lattice action is not invariant under gauge
transformations:
$S^\pr(U) \not= S^\pr(U^g)$, where
$U$ is the lattice gauge field and
\be
U^g_{\m x}=g_x U_{\m x}g^\dg_{x+\mh}    \label{UG}
\ee
is the gauge transform of $U$. We shall restrict ourselves to compact gauge
fields here, with $U$ in a gauge group $G$.
For instance, in the case of lattice actions for chiral gauge theories
the gauge symmetry breaking  resides in the
action for the fermions and in gauge variant counter terms for the
gauge field that have to be added.

In the path integral one could integrate over the full gauge
field configuration space, which includes the longitudinal modes
(corresponding to gauge transformations) of $U$. Since the action
$S^\pr$ is not gauge invariant, these modes do not decouple and
in proposals for chiral gauge theories on the lattice, they are found
to destroy the chiralness of the model \cite{SmSw92,BoSm93,GoJa93}.
Alternatively, as in the Rome proposal \cite{Rome}, the longitudinal modes
can be removed from the path integral by fixing the gauge to a smooth
transversal gauge.

Gauge fixing is usually implemented by inserting a delta function in
the path integral that has its support on a gauge section specified
by some suitable gauge condition. For the moment we shall write this
gauge condition as a constraint of the form $F_{GF}(U^g)=0$.
An example of such a condition is the local Landau condition,
$\del_\m A_\m=0$, which on the lattice takes the form,
\be
 F^L_x(U^g) = \sum_\m \IM [ U^g_{\m x} - U^g_{\m x-\mh}] = 0.  \label{LANDAU}
\ee
To specify an (almost) unique solution to this condition, one should
find the $g$ such that the function
\be
H(U^g) = \sum_{x \m} \RE\, \Tr U^g_{\m x}. \label{LANDF}
\ee
achieves its absolute maximum.
This shows that in this gauge the average value of the link field,
$\av{\Tr U} = \av{H(U)}/H(1)$, is as close to one as possible.
We shall call gauge fixed configurations that have a large value
of the average link, `smooth'. Note that the Landau gauge does not
fix the global gauge invariance.

By including the  factor $\d(F_{GF}(U))$ in the path integral,
$\int DU \d(F_{GF}(U))\exp S(U)$,
the gauge invariance is removed.  Now the model resembles any other
non-gauge invariant model, in which the saddle point configurations carry
the same action as in the continuum model, up to $O(a)$ effects,
and nonsmooth configurations contributing to the path integral account
for quantum effects.
However, as is well known,
most constraints require a Jacobian factor $\D_{GF}$, such that
$\int Dg \D_{GF}(U)\d(F_{GF}(U^g)) = 1$. Unfortunately, such a
Fadeev-Popov Jacobian is difficult to handle numerically.

Our approach avoids imposing a constraint on the path integral, but
guarantees that only the selected maximally smooth configurations
$U^g$ contribute to
the path integral by defining a new, almost gauge invariant action $S$ as
\be
   S(U) = S'(U^{g(U)}), \Sp F_{GF}(U^{g(U)})=0.
	                 \label{SNEW}
\ee
$S^\pr$ is the gauge variant lattice action that reduces in the
classical continuum limit to the gauge invariant continuum action we
wish to regularize and
$U^{g(U)}$ is the gauge fixed gauge field, $F_{GF}(U^{g(U)})=0$, which
is unique up to a global gauge transformation.
The action $S(U)$ is well-defined for all $U$, provided that the gauge
condition $F_{GF}$ specifies the gauge transformation $g(U)$
uniquely for all $U$. This means that the
gauge condition should not suffer from Gribov ambiguities.

Typically the gauge condition does not fix the global invariance and to
define $g(U)$ unambiguously, the condition has to be amended with an
(arbitrary) prescrition, such as $g(U)_{x_0}=1$ on a given site $x_0$.
This implies that under a gauge transformation $h$ on $U$,
$g(U)$ transforms as,
\be
  g(U^h)_x = h_0g(U)_xh_x^\dg,
\ee
with $h_0=h_{x_0}$ for the prescription mentioned above.
This transformation rule implies that the action (\ref{SNEW}) is gauge
invariant, up to the global transformation $h_0$,
\be
S(U^h) = S'((U^h)^{g(U^h)})=S'((U^h)^{h_0g(U) h^\dg}) = S'(U^{h_0g(U)})
       = S(U^{h_0}).
\ee
For abelian gauge groups $U^{h_0}=U$ and then the action is actually
gauge invariant.

Since the action is not invariant (for non-abelian gauge groups) under
constant gauge transformations, the mode $h_0$ appears as a new degree
of freedom in the path integral.
However, such a global mode should not affect the physics of the model
and it must be emphasized that the dangerous local gauge degrees
of freedom, the longitudinal modes of the gauge field, decouple because
of the `almost' gauge invariance of the action.

It is important to note that we use the
action $S$ in the path integral {\em without} restricting
the integration to a particular gauge section, as would be
the case with the usual Fadeev-Popov $\d$-function method. Therefore
it avoids the complications of a Jacobian factor. It also implies,
of course, that the integration along the
gauge orbits is not damped. In a continuum model, it would then still
require gauge fixing to make this path integral well-defined, but on the
lattice this is not necessary. If we would, however, add gauge fixing
and ghost terms (as is required in perturbation theory), then our approach
would correspond to the Rome approach \cite{Rome},
assuming we impose exactly the
same gauge condition $F_{GF}$ as used to define the action (\ref{SNEW}).

The new action (\ref{SNEW}) contains the gauge
transformation $g(U)$ which is a nonlocal function of the gauge field.
We shall skip over possible formal complications of defining a transfer
matrix and a Hilbert space corresponding to this nonlocal action.
Also on a pragmatic level, however, one might fear
that the nonlocality of $g(U)$ makes the action (\ref{SNEW})
unsuitable for numerical simulations.
Fortunately this turns out not to be the case.

\section{Algorithms for a numerical simulation with $S$}

A straightforward Metropolis algorithm with the action (\ref{SNEW})
would be cumbersome, because for each link-update
$U_{\m x} \ra U^\pr_{\m x}$
the new action must be computed, which needs the $g(U^\pr)$. This
can be computed from the condition $F_{GF}((U^\pr)^{g^\pr})=0$, but
this is typically a time consuming computation.

Alternatively, one can aim for a hybrid Monte Carlo (HMC) algorithm.
To find such an algorithm, we proceed as usual
\cite{Kenn85} and introduce the momenta $\p$
corresponding to the gauge potential, and the Hamiltonian,
\be
  H =  \sum_{x \m a} \half (\p_{\m x}^a)^2 + S(U).    \label{HAM}
\ee
The gauge fields and momenta evolve in time, such that the Hamiltonian
remains constant. Choosing the convenient evolution equation\footnote{
The $T^a$ are the hermitian generators of the algebra of the gauge
group,  $[T^a,T^b]  =  if^{ab}_cT^c$ and $\Tr (T^aT^b) = \d^{ab}/2$.}
\be
     \dt U_{\m x}  = i\p_{\m x}^aT^aU_{\m x}, \label{EOMU}
\ee
for the gauge field $U$, we can find the equation for the momentum from
the condition that $\dt H = 0$. This gives
\be
     \dt \p_{\m x}^a = F_{\m x}^a(U) = \cF_{\m x}^a(U) + \cG_{\m x}^a(U),
                   \label{EOMP}
\ee
with the two terms in the force defined by
\bea
   \cF_{\m x}^a(U) &\! =\! & \IM\left[
   \frac{\del S(U,U^\dg,g,g^\dg)}{\del U^{bc}_{\m x}}
                               \left(T^a U_{\m x}\right)^{bc}
                        \right],  \label{FU}\\
   \cG_{\m x}^a(U) &\! =\! & \IM\left[
      \frac{\del S(U,U^\dg,g,g^\dg)}{\del g^{bc}_y}
      \frac{\del g^{bc}_y}{\del U^{de}_{\m x}}\left(T^a U_{\m x}\right)^{de}
  +  \frac{\del S(U,U^\dg,g,g^\dg)}{\del g^{bc*}_y}
     \frac{\del g^{bc*}_y}{\del U^{de}_{\m x}}\left(T^a U_{\m x}\right)^{de}
           \right],  \label{FG}
\eea
where we have suppressed the summations over the group indices.
The first part of the force, $\cF$, is the usual term coming from the
direct $U$ dependence of the action; the second term, $\cG$, comes from
the implicit $U$ dependence through the gauge transformations $g(U)$.
The partial derivatives $\del g_y(U)/\del U_{\m x}$ and
$\del g_y(U)^*/\del U_{\m x}$ can be solved from the gauge
condition, as is shown in the next section.
We could also try to solve the time dependence of $g(U)$ from the
equation
\be
  \dt g_x  = \sum_{y \m a}
  i\p^a_{\m y}\left[\frac{\del g_x}{\del U_{\m y}}\left(T^a U_{\m y}\right)
      -    \frac{\del g_x}{\del U^\dg_{\m y}}\left(U^\dg_{\m y} T^a\right)
	     \right].
                                    \label{EOMG}
\ee

If the equations (\ref{EOMU}), (\ref{EOMP}) and (\ref{EOMG}) could be
solved exactly, the
Hamiltonian would be constant in time and the field $U^g$ would always
satisfy the gauge condition $F_{GF}(U^g)=0$ (at least in a region
where the gauge section has no singular points).
The inevitable discretization of eq. (\ref{EOMU})
spoils conservation of energy, but this is remedied by including the usual
accept/reject step. However, discretization errors in (\ref{EOMG}) also
affect $g(U)$, such that the gauge fixed field $U^{g(U)}$ drifts
away form the gauge section. Also when encountering a singularity in the
gauge condition (a Gribov horizon), the solution of (\ref{EOMG})
will no longer give the correct $g(U)$.  This interferes with the
computation of $\cG(U)$, because $\del g/\del U$ can only be computed from
the constraint
under the assumption that $U^{g(U)}$ satisfies the gauge condition exactly.

Since we cannot rely on solving (\ref{EOMG}) to obtain $g(U)$,
we include an exact evaluation of the required gauge transformation,
denoted symbolically by $g(U) = F_{GF}^{-1}(U)$, in each time step.
This leads to the discretized (leap frog) update over one time step
$dt$,
\bea
  \p_\m^a(t+\half dt) & = & \p_\m^a(t) + \half F^a_\m(U(t),g(t)) dt,
                                     \label{PIUP1}  \\
   U_\m^a(t + dt)     & = & \exp{(iT^a\p^a_\m(t + \half dt)dt)}U_\m^a(t),\\
                                     \label{AUP}
   g(t+dt)            & = & F_{GF}^{-1}(U(t+dt))           ,      \\
                                     \label{GUP}
  \p_\m^a(t + dt)     & = & \p_\m^a(t+\half dt)
                        + \half F^a_\m(U(t+dt),g(t+dt)) dt.
                                     \label{PIUP2}
\eea
By iterating this process, one can make large update changes in the
gauge field, with a large acceptance rate if the accumulating
discretization errors are not too large.
This algorithm satisfies  detailed balance, provided that the update
rule (\ref{PIUP1}--\ref{PIUP2}) is reversible and area preserving.

To check reversibility, we must show that a subsequent update of $U(t+dt)$
and $-\p(t+dt)$ with the rule (\ref{PIUP1}--\ref{PIUP2}) leads to
$U(t+2dt) = U(t)$ and $\p(t+2dt)=-\p(t)$.  This is easy to verify,
provided that the gauge transformation is uniquely fixed for any $U(t)$,
which we assume to be the case. Area preservation of the map follows
from the structure of the leap-frog equation. Our scheme only deviates
from the standard algorithm by the extra $g$ dependence of the force
term. But this only modifys the $U$ dependence of $F$, it does not lead
to additional $U$ dependence in the update rule (\ref{AUP}) for $U$, nor
does it introduce $\p$ dependence in the update rules (\ref{PIUP1})
and (\ref{PIUP2}) for $\pi$. Therefore one can check that each of the
three maps $(U(t),\pi(t)) \ra (U(t),\pi(t+\half dt))$,
$(U(t),\pi(t +\half dt)) \ra (U(t+ dt),\pi(t+\half dt))$ and
$(U(t+dt),\pi(t +\half dt)) \ra (U(t+ dt),\pi(t+ dt))$ leaves the
(Haar) measure $dUd\p$ invariant.

The reversibility of the update rule is seen to depend crucially on the
uniqueness of the gauge section. One runs into problems when
using the Landau condition (\ref{LANDAU}) as gauge condition $F_{GF}(U^g)$,
because this condition has many solutions $g$ for a given $U$.
 Even when it is supplemented with the
condition that the solution must be an absolute maximum of the function
$H$ defined in eq. (\ref{LANDF})
this gauge condition cannot be used in
practice, because it is not possible to find the absolute maximum of
$H$ in a reasonable amount of computer time. Therefore we shall choose
a different gauge condition from which $g$ can be unambiguously computed
in practice for (almost) any $U$.

\section{Laplacian gauge fixing condition}

In ref. \cite{ViWi92} a gauge fixing condition was introduced, which was
argued to produce smooth gauge fixed gauge fields, and which can be found
unambiguously for (almost) any
gauge field $U$.  The gauge transformation $g$ to put $U$ in this
`Laplacian' gauge, is
computed from the eigenfunctions of the covariant Laplacian,
\be
\D(U) f^s =  -U_{\m x} f^s_{x+\mh} - U_{\m x-\mh}^\dg f^s_{x-\mh} =
              \l^s f^s_x,   \label{LAPL}
\ee
where we discarded a term proportional to $\un$ and suppress the gauge
field indices, as before. For gauge group $G=SU(N)$, the gauge transformations
are computed from the $N$ eigenfunctions with smallest eigenvalues.
These functions define an $N\times N$ matrix field $M$,
\be
  M^{ab}_x = f^{ab}_x,
\ee
where $a$ is the $SU(N)$ index of the eigenfunctions and $b$ labels the
$N$ selected eigenfunctions with eigenvalues $\l^b$ in increasing order,
 $\l^1\leq \l^2, \cdots,\leq \l^N$.
The gauge transformation $g\in SU(N)$ is obtained from $M$ as
\be
  g = M P^{-1}(\Det(M^{-1}P))^{1/N}, \;\; P^2 = M^\dg M. \label{GLAP}
\ee

As discussed in more detail in ref. \cite{ViWi92} this prescription is
unambiguous,
except when the $N^{th}$ eigenvalue is degenerate with the $(N+1)^{th}$,
or when any $M_x$ has one or more zero modes.  The global phases of
the eigenfunctions are arbitrary, which implies that $g$ is only
defined up to a global factor\footnote{For gauge froup $U(N)$ this is
a global $U(1)^N$ transformation, for $SU(N)$ it is $U(1)^{(N-1)}$;
for $SU(2)$ it is an $SU(2)$ transformation, because the eigenvalue
spectrum  has an exact two-fold degeneracy \cite{ViWi92}.}.
Configurations which give rise to a coincidental degeneracy
of the $N^{th}$ and $(N+1)^{th}$ eigenvalue, have measure zero in the full
configuration space.
However, the sequence of configurations produced by a HMC algorithm could
encounter these singular points in the continuous time limit $dt\ra 0$.
With finite $dt$ the chance of hitting a singular configurations would
then still be zero since these points will fall in between two
subsequent time steps. In an actual implementation, of course,
the probability of
encountering a singular configuration depends on the numerical accuracy
with which the eigenvalues and eigenfunctions can be computed.
Further more, there is the possibility that in the region of
configuration space traced out by the HMC trajectories, there is no
actual level crossing, but only `avoided' crossing. The occurence of
(avoided) level crossing and the numerical difficulties it may entail
should become clear in an actual simulation.

In the update step for the momenta we must compute the extra term in
the force, $\cG$, which contains
the term $\del g/\del U$, cf. eq. (\ref{FG}). We consider here
gauge group $U(1)$ for simplicity, for $SU(N)$ the computation is
more complicated by the additional group structure.
For gauge group $U(1)$ we only need one eigenfunction, and the
gauge transformation is defined as,
\be
  g_x = f^{0*}_x|f^0_x|^{-1},
\ee
with $f^0$ the eigenfunction of the Laplacian with smallest
eigenvalue.
First we compute the change of $f^0$  under a small update
of the gauge field $U \ra e^{idt \p_{\m}}U = U + \d U$.
This update of $U$ changes the Laplacian (\ref{LAPL}) to $\D + \d\D$ with
\be
  (\d\D f^s)_x = -\sum_\m (
     \d U_{\m x}f^s_{x+\mh} + \d U^*_{\m x-\mh}f^s_{x-\mh} ).
\ee

Perturbation theory gives the
corresponding change in the eigenfunction as,
\be
\d f^0_y =  \sum_x \sum_{t\not=0}
        \frac{f^{t}_yf_x^{t*}}{\l^0 - \l^t} (\d\D f^0)_x ,
\ee
where $t$ runs over all (normalized) eigenmodes $t\not= 0$.
In order to avoid this sum involving all eigenfunctions of
the Laplacian, we note that
\be
 \sum_{t\not=0} \frac{f^t f^{t*}}{\l^t - \l^0}
 =  (\D - \l^0)_{\perp 0}^{-1},
\ee
where the inverse of $(\D - \l^0)$ must be taken on the subspace
orthogonal to $f^0$. This can be computed numerically with an iterative
(e.g. conjugate gradient) algorithm, for the matrix restricted to this
subspace,
\be
 (\Db h)_x := ((\D - \l^0)_{\perp 0}h)_x =
             (\D_{xy} -\l^0\d_{xy})[h_x - (\sum_zf_z^{0*}h_z) f^0_x],
\ee
with $h$ an arbitrary vector.

With these definitions we have,
\be
 \d f^0_y = -\sum_z (\Db)_{yz}^{-1} (\d\D f^0)_z.
\ee
Untill now we have followed the usual perturbative convention for the
phase of the perturbed eigenfunction given by the condition that
$\sum_z f^{0*}_z\d f_z = 0$. In our HMC scheme we want to impose the
diffent condition that $\d g_{y_0} = 0$.  This can be achieved by adding
$\d\a f^0_y$ with properly choosen $\d\a$.
For the  variation of the gauge transformation we then find,
\be
 g_y^*\d g_y = \half (g_y^* |f^0_y|^{-1}\d f^{0*}_y - c.c.) - (y\ra y_0),
\ee
where $c.c.$ is the complex conjugate and the term with $y\ra y_0$
is a global phase shift that ensures $\d g_{y_0}=0$.
This leads to
\bea
   g_y^*\frac{\del g_y}{\del U_{\m x}}U_{\m x} & = &
 \half |f^0_y|^{-1}\left( g^*_y(\Db)^{-1*}_{y x+\mh} U_{\m x}f^{0*}_x
 -      g_y(\Db)^{-1}_{y x} U_{\m x}f^0_{x+\mh}\right)
 - (y\ra y_0),\nnn
   g_y\frac{\del g^*_y}{\del U_{\m x}}U_{\m x} & = &
 \half |f^0_y|^{-1}\left( g_y(\Db)^{-1}_{y x} U_{\m x}f^0_{x+\mh}
 -         g^*_y(\Db)^{-1*}_{y x+\mh} U_{\m x}f^{0*}_x \right)
 - (y\ra y_0),
\eea
from which we can compute the force term $\cG$ as,
\bea
  \cG_{\m x}(U) & = & 2\RE \left( f^{0*}_xU_{\m x} w_{x+\mh}
                            - f^{0*}_{x+\mh}U^*_{\m x} w_x \right),\nnn
   w_x & = & \sum_y(\Db)^{-1}_{xy} |f^0_y|^{-1} g^*_y
   \IM\left(g_y\frac{\del S}{\del g_y}
   -\d_{y,y_0}\sum_zg_z\frac{\del S}{\del g_z}\right),
\eea
where we have also used the hermiticity of $\Db$.

For non-abelian gauge groups $SU(N)$, the formula is more involved,
but it has the same structure. The important feature is that the
computation of the force term requires an inversion of the hermitian matrix
$\Db$ to compute the vector $w$ ($N$ inversions are needed for $SU(N)$).
This makes the algorithm time consuming, but not much worse than is the case
in simulations with dynamical fermions, where a non-hermitian matrix must
be inverted each time step.

At this point two further remarks are in order.
First, we have noted above that as a function of the HMC time, there
may  be level crossings between the  $N^{th}$ or $(N+1)^{th}$ eigenvalues of
the Laplacian. At these points the linear perturbation estimate
computed above, will in general, of course, not be an adequate estimate
for the change in $g$ and the $U$ dependence of the Hamiltonian is not
continuous (with avoided crossing the Hamiltonian is continuous but may
be rapidly changing). As a consequence the change in the Hamiltonian
(\ref{HAM}) can be larger than the usual $O(dt^2)$, for such a time step.
This depends on the magnitude of the gauge symmetry breaking, since
only such terms are sensitive to a (large) change of $g$. As a result,
it could lead to small acceptance rates when level crossing is
encountered along a HMC trajectory. However, the algorithm
preserves detailed balance, also when encountering such a level crossing
horizon in configuration space, because detailed balance
only requires that $g$ is a unique function of $U$; it is not required
that it is a continuous function of the HMC time $t$.

Second, we have assumed that the Laplacian gauge is a smooth gauge
and in ref. \cite{ViWi92} this was argued to be plausible, because
the eigenfunctions of the Laplacian with small eigenvalues should
be smooth functions, up to the gauge transformation $g$. However, it
is important to verify this numerically, because only
if the path integral is dominated by sufficiently smooth configurations,
we may hope that there exists a scaling region in which
the dangerous longitudinal gauge degrees of freedom do not affect the
physics.

\begin{figure} [ttt]
 \centerline{ \epsfysize= 9.5cm   \epsfbox{ulink.ps}
 }
\vspace{-18mm}
\caption{  Average link $\av{U}$ as a function of the inverse gauge
coupling $\b$, after standard Landau gauge fixing
(solid line), Laplacian gauge fixing (dashed line) and Laplacian
followed by Landau gauge fixing (dotted line), for U(1) gauge
fields in two dimensions on a lattice with volume $20^2$.
}
\end{figure}

As a first step in an investigation of Laplacian gauge fixing, we have applied
it to compact $U(1)$ gauge fields in two dimensions. As a measure of the
smoothness of the gauge fixed configuration, we use the value of
the average link. If $\av{U_{\m x}}$ is close to one, the configuration
is smooth. In fig.~1 we have plotted this average link as a function of
the inverse gauge coupling $\b$. The full line is the result after
Laplacian gauge fixing, the dashed line is the result after standard
Landau fixing. Here we use a checker board relaxation algorithm to
maximize the function (\ref{LANDF}). One sees, that the Laplacian
gauge fixing leads to smoother gauge fixed configurations for
$\b \apgt 3$, which roughly corresponds to the scaling region of this model.
Only for small
$\b$ the usual Landau fixing is smoother, but even there the difference
is not dramatic. The third curve (dots) is obtained by applying
Landau gauge fixing {\em after} putting the gauge field in the Laplacian
gauge.  Here we consistently find that the subsequent Landau cooling
increases the average link to a somewhat larger value. This indicates
that the standard Landau gauge fixing typically is unable to find the
absolute maximum of (\ref{LANDF}).  This Gribov ambiguity in standard Landau
gauge fixing has previously been observed in $U(1)$ and $SU(N)$ models,
see e.g. refs. \cite{GriLat}.

\section{A simple example}

To illustrate our approach, we consider a simple toy model, consisting
of a periodic lattice with two sites and link fields $U_i=e^{i\q_i}$,
$i=1,2$. For the action we choose
\be
 S^{toy} = \b \cos(\q_1 + \q_2) + \k \sin(\q_1 - \q_2).
                       \label{STOY}
\ee
For $\k=0$ this model has the gauge invariance $U_1 \ra g_1U_1g_2^*$,
$U_2 \ra g_2U_2g_1^*$. With the global phase fixed by the choice $g_2=1$
and writing $g_1=e^{i\g}$,
this amounts to the transformation rule,
\be
 \q^\g_1 = \q_1 + \g, \;\; \q^\g_2 = \q_2 - \g;
\ee
the $\k$-term breaks the gauge symmetry.
The gauge orbits in this model are the lines of constant $\q_1 + \q_2$.
By gauge fixing we should select precisely one point on each of these
orbits.

Landau gauge fixing in this model
would be defined by maximizing the function $H$ of eq. (\ref{LANDF}),
which in this model is
\be
  H = \cos(\q_1+\g) + \cos(\q_2-\g).  \label{LANTOY}
\ee
This has the solution $\g =-(\q_1 - \q_2)/2$, or equivalently
$\q^\g_1 = \q^\g_2$, with $-\p<(\q^\g_1 + \q^\g_2)<\p$ (mod $2\p$).
Outside this interval, the line $\q^\g_1=\q^\g_2$ is still
an extremum of (\ref{LANTOY}), but now it is a minimum.
At the `horizon' $\q_1 + \q_2 = \pm \p$, $H$ is zero for all values of $\g$
and the gauge section is undetermined on this orbit.

In this simple model
we could also use the  Landau gauge fixing condition in our HMC algorithm,
because the prescription to choose the absolute maximum of $H$
uniquely specifies the gauge, except for the single orbit at the horizon
$\q_1+\q_2=\pm \p$, and a straightforward cooling algorithm would easily
find this absolute maximum, also when the starting point would be just
a distance $\e$ across the horizon, for instance at $\q_1=\q_2$
with $\q_1=\p/2 + \e$.
However, in a realistic model, such a large shift to a different global
maximum would be impossible to find in practice.

For Laplacian gauge fixing, we first have to solve the eigenvalue
equation for the Laplacian on this two site lattice. The Laplacian
is a two by two matrix,
\be
  \D(U) = -\left( \begin{array}{cc}
                       0     &     U_1 + U_2^* \\
                  U_1^* + U_2&     0
                 \end{array} \right).
\ee
The eigenfunctions and eigenvalues are
\be
  f_{\pm} = \left( \begin{array}{c}
                    \mp e^{i\f} \\ 1
                    \end{array} \right) ,\;\;
  \l_{\pm} = \pm | \cos((\q_1+\q_2)/2) |,
\ee
with the phase $\f$ defined by
\be
  \f = (\q_1-\q_2)/2 + \p \Q(-\cos((\q_1+\q_2)/2)).
\ee
The  step function $\Q$ in the phase $\f$ of the
eigenfunction provides the phase shift over $\p$ induced by the level
crossing that occurs at the horizon $\q_1+\q_2 = \p$ (mod $2\p$).

The gauge transformation follows from the eigenfunction
with the smallest eigenvalue, $f_-$, which gives $g_1 = e^{-i\f}$,
$g_2=1$.
This is exactly the same gauge transformation that defines the
Landau gauge section. Also the Gribov horizon of the Landau gauge
is seen to correspond precisely to the level-crossing horizon in
the Laplacian gauge.

\begin{table} \begin{center}
\begin{tabular}{|c|c||c|c||c|c|}
  \hline
 $\b$   & $\k$ & $S_1$, HMC&$S_1$, exact&$S_2$, HMC &$S_2$, exact \\
  \hline \hline
  1     &   0      & 0.4456(25)&0.44639 & 0.4456(25)& 0.44639 \\
        &   1      & 0.4467(29)&0.44639 & 0.5617(25)& 0.56169 \\
        &   2      & 0.4464(30)&0.44639 & 0.6438(24)& 0.64276 \\
        &   3      & 0.4468(20)&0.44639 & 0.6978(26)& 0.70028 \\
  \hline
  3     &   1      & 0.8101(14)&0.80999 & 0.8288(15)& 0.83017 \\
        &   2      & 0.8127(25)&0.80999 & 0.8451(13)& 0.84580 \\
        &   3      & 0.8095(44)&0.80999 & 0.8580(11)& 0.85837 \\
  \hline
\end{tabular}
\end{center}
\caption{ results for \protect{\av{\cos(\q_1 + \q_2)}} in the toy model
with gauge breaking term $S_1=\k \sin(\q_1 - \q_2)$ and
$S_2=\k \sin(\q_1)$; the columns with `HMC' give the results obtained with the
HMC algorithm, `exact' gives the exact results. }
\end{table}

As a test of the HMC algorithm (\ref{PIUP1}--\ref{PIUP2}), we have
implemented it for this toy model. Here we can compute expectation values
like $\av{\cos(\q_1 + \q_2)}$ both analytically and with our
HMC algorithm. In table 1 we give a few of these results obtained
for various values of the $\k$ in front of the gauge symmetry breaking
term in (\ref{STOY}). Since this term is zero on the gauge section,
the results should not depend on it, as is seen to be the case for
the results in the first column  (HMC results) and second column
(exact result).  The last two columns contain results obtained with
a different gauge symmetry breaking term, $\k \sin(\q_1)$. This term
is nonzero on the gauge section and now the results are seen to
depend on $\k$, but the HMC results are fully consistent with the
exact answers.

\section{Discussion}

The HMC algorithm defined in eqs. (\ref{PIUP1}--\ref{PIUP2}) for the
first time provides
a manageable method to simulate models with broken gauge invariance,
such as lattice models for chiral gauge theories,
in which the action is always computed on a smooth gauge section.
Starting from a gauge variant action, we define a new action in eq.
(\ref{SNEW}),
which is made (almost) gauge invariant using
the gauge transformation $g(U)$ that defines the Laplacian gauge of the
configuration $U$. For abelian gauge groups the new action is gauge
invariant.
We use the Laplacian gauge because it presently is the only smooth,
Landau-like, gauge which can be computed unambiguously in a reasonable
amount of computer time.

In simulations with this new action
the gauge need not be fixed and hence no ghost term is required.
In this important aspect our approach differs from the Rome approach
\cite{Rome} which includes gauge
fixing and ghost terms in its lattice action; it also differs from
previous gauge invariant approaches which use an additional scalar field
to restore the gauge invariance, whereas we use the field $g(U)$ which
is a function of the gauge field $U$.

The HMC algorithm is numerically demanding, because the gauge fixing
field $g(U)$ has to be computed for each gauge field on the trajectory.
This requires the computation of the $N$ eigenfunctions with smallest
eigenvalues of the gauge invariant Laplacian (for gauge group $SU(N)$).
The momentum update requires $N$ conjugate gradient inversions
for each time step.  This clearly makes the algorithm very time consuming,
but not much worse than a simulation with dynamical fermions.
The eigenfunctions can be computed using a Lanczos algorithm, combined
with inverse iteration.
This computation should not be particularly hard, except
perhaps in the vicinity of level crossing horizons, which  are the Gribov
horizons of the Laplacian gauge.
Also the acceptance rate of the HMC algorithm could deteriorate when
crossing such a level crossing horizon.
Preliminary results for $SU(2)$ Laplacian gauge fixing on a lattice
with volumes $V=8^4$ and $16^4$, indicate that
for typical configurations the eigenvalues and eigenfunctions can
be computed to a precision $\d\l < 10^{-10}$ and $\d f<10^{-8}$,
with $\d\l = |\l^{exact}-\l^{found}|$ and
$\d f = (\sum_{x a} |f^{a,exact}_x-f^{a,found}_x|^2/V)^{1/2}$.
These results also showed that the Laplacian gauge can be implemented
efficiently and that it leads to smooth
gauge fixed configurations, with link expectation values which are very
close to those in the Landau gauge.

In this letter we have shown that a model in which gauge invariance is
broken in the lattice action, can be made gauge invariant (up to a
global gauge transformation) and we have discussed a HMC
algorithm for simulations with this new class of actions.
To apply this method to chiral gauge models, one could use e.g. the
fermion methods considered in refs. \cite{SmSw92,BoSm93,GoJa93} and
replace the scalar field $V$ of these models by $g(U)^\dg$
(or by $g(U)$ in the conventions of ref. \cite{GoJa93}).
An actual investigation of the resulting models, should e.g. include
a study of the global and local anomalies. This is, however,
postponed to future work.\\[2mm]\noindent

I would like to thank W. Bock, M. Golterman, J. Hetrick,
K. Jansen, J. Kuti and P. van Baal for illuminating discussions.
This work is supported by the DOE under grant DE-FG03-91ER40546
and by the TNLRC under grant RGFY93-206.


\begin{thebibliography}{99}
\bibitem{Petc93} D.N. Petcher, Nucl. Phys. B (Proc. Suppl.) 30 (1993) 50.
\bibitem{SmSw92} M.F.L Golterman, D.N. Petcher and J.Smit,
                 Nucl. Phys. B370 (1992) 51.
\bibitem{BoSm93} W. Bock, J. Smit and J.C. Vink, preprint
                 ITFA 93-18, UCSD/PTH 93-15, hep-lat 9308009.
\bibitem{GoJa93} M.F.L Golterman, K. Jansen, D.N. Petcher and J.C. Vink,
                preprint UCSD/PTH 93-28, Wash. U. 93-60, hep-lat 9309015.
\bibitem{Rome}   A. Borelli, L. Maiani, G.C. Rossi, R. Sisto and
                 M. Testa, Phys. Lett. B221 (1989) 360; Nucl. Phys.
                 B333 (1990) 355;
	         L. Maiani, Nucl. Phys. B (Proc. Suppl.) 29B,C (1992) 33.
\bibitem{Grib78} V. Gribov, Nucl. Phys. B139 (1978) 1.
\bibitem{GriLat} A. Nakamura and M. Plewnia, Phys. Lett. 255B (1991) 274.
                 Ph. de Forcrand, J.E. Hetrick, A. Nakamura and
                 M. Plewnia, Nucl. Phys. B(Proc. Suppl.) 20 (1991) 194.
                 E. Marinari, C. Parrinello, R. Ricci, Nucl. Phys. B362
                 (1991) 487.
\bibitem{Kenn85} S. Duane, A.D. Kennedy, B.J. Pendleton and
                 D.~Roweth, Phys. Lett. B195 (1987) 216.
\bibitem{ViWi92} J.C. Vink and U.-J. Wiese, Phys. Lett. B289 (1992) 122.
\end{thebibliography}
\end{document}